\expandafter\ifx\csname natexlab\endcsname\relax\fi

\documentclass[numberedappendix]{emulateapj}
\slugcomment{{\sc Accepted to ApJ:} August 18th, 2018} 
\usepackage{graphicx}
\usepackage{epstopdf}
\usepackage{color}
\usepackage{amsmath}
\usepackage{mathtools}
\usepackage{hyperref}
\usepackage{perpage}
\MakePerPage{footnote}
\def\a4{\hsize 17.0cm \vsize 25.cm}

\defcitealias{MRN1977}{MRN} 
\defcitealias{MartinezGonzalezetal2016}{MTS16}
\defcitealias{MartinezGonzalezetal2017}{MWP17}
\defcitealias{Kroupa2001}{Kroupa}
\defcitealias{Bocchioetal2016}{B16}
\defcitealias{Ferraraetal2016}{Ferrara et al.}
\shorttitle{}

\shorttitle{Pyroclastic Blowout}
\shortauthors{Mart\'inez-Gonz\'alez  et al.}

\begin{document}

\title{Pyroclastic Blowout: Dust Survival in Isolated versus Clustered Supernovae}

\author{Sergio Mart\'inez-Gonz\'alez\altaffilmark{1} ,
 Richard W\" unsch\altaffilmark{1},
 Jan Palou\v s\altaffilmark{1},
 Casiana Mu\~{n}oz-Tu\~{n}\'on\altaffilmark{2,3}, 
 Sergiy Silich\altaffilmark{4},
 Guillermo Tenorio-Tagle\altaffilmark{4}}
 
\altaffiltext{1}{Astronomical Institute, Czech Academy of Sciences,  Bo\v{c}n\'\i\ II 1401/1, 141 00 Praha 4, Czech Republic; martinez@asu.cas.cz}
\altaffiltext{2}{Instituto de Astrof\'isica de Canarias, E 38200 La Laguna, Tenerife, Spain}
\altaffiltext{3}{Departamento de Astrof\'iısica, Universidad de La Laguna, E 38205 La Laguna, Tenerife, Spain}
\altaffiltext{4}{Instituto Nacional de Astrof\'\i sica, \'Optica y Electr\'onica, AP 51, 72000 Puebla, M\'exico}

\begin{abstract}
Following the current debate on the fate of SN-condensed dust grains, here we explore by means of three-dimensional hydrodynamical 
simulations the interaction of dusty supernova remnants (SNRs) with the shocked winds of neighboring massive stars within young massive stellar
clusters (SSCs). As a comparison, we have also explored the evolution of supernova remnants in the diffuse ISM with constant density. Since 
the hydrodynamics of SNRs is intimately related to the properties of their immediate environment, the lifecycle of dust grains in SNRs within SSCs
is radically different from that in the diffuse ISM. Moreover, off-centered SNRs evolving in the steep density gradient 
established due to a star cluster wind experience a blowout phase: shell fragmentation due to protruding Rayleigh-Taylor instabilities
and the venting of SN ejecta. Our main finding is that clustered SN explosions will cause a net increase in the amount of dust in the surroundings of young 
massive stellar clusters. Our analysis considers the multiple dust processing resulting from the passage of the SN reverse shock, including its reflection at 
the SNR's center, the injection of shocked stellar winds within the respective remnant's volume and the effect of secondary forward shocks produced in sequential 
SN explosions. In the simulations, we have on-the-fly calculated the rates of thermal sputtering and dust-induced radiative cooling provided an initial distribution 
of grain sizes and dust content. Fast-moving elongated dusty SN ejecta resemble mushroom clouds violently ascending in a stratified atmosphere after 
volcanic super-eruptions, where the pyroclasts carried by the clouds are wind-driven and eventually accumulate into the vast surroundings. 
\end{abstract}


\section{Introduction}
Dust nucleation and subsequent growth in the ejecta of core-collapse supernovae (SNe) are nowadays recognized as highly efficient and rapid 
processes \citep{Cernuschietal1967,TodiniandFerrara2001}. To put it into perspective, it has been estimated that up to $0.4$ to $1.1$ M$_{\odot}$ of dust 
can be formed per SN, as in the case of several nearby supernova remnants (SNRs) like SN 1987A \citep{Indebetouwetal2014,Matsuuraetal2014}, 
Cassiopeia A \citep{DeLoozeetal2017,Bevanetal2016} and G54.1+0.3 \citep{Temimetal2017,Rhoetal2017}. Modeling dust injection by SNe is key to 
understand the process of dust enrichment of galaxies at all cosmic epochs \citep[needed to explain the otherwise puzzling amounts of dust observed in 
galaxies during cosmic reionization,][]{Laporteetal2017}. However, the fate of the dust grains formed in SN ejecta remains a matter of intense debate 
as they must be processed by the shocks that are formed from the interaction of freely-expanding SNRs with their ambient medium: a 
forward shock which sweeps the surrounding ISM, and a reverse shock which moves through the SN ejecta. The former might lead
to the destruction of the pre-existent dust in the surroundings, while the latter might be capable of inducing the destruction of a large 
fraction of the ejecta dust primarily via thermal sputtering. Notwithstanding, while there are thoughtful arguments that put into question the 
efficiency of dust destruction by SN shocks \citep[e.g.][]{Jonesetal2011,Ferraraetal2016}, recent chemical and dynamical evolution models have favored the notion that only a minor 
fraction of this dust ($\lesssim 10\%$ by mass) will be able to survive the passage of the respective SN reverse shock 
\citep[see e.g.][]{Nozawaetal2007,Micelottaetal2016,Bocchioetal2016}. For instance, \citet{Bocchioetal2016} (hereafter B16) followed the dust mass, chemical 
composition and grain size distribution of several dust species in SN ejecta taking into account the grain dynamics during the 
free-expansion, Sedov-Taylor and snowplough phases of the evolution of SNRs. The influence of turbulent ISM magnetic fields was studied in detail by 
\citet{Fryetal2018}, whom, remarking the importance of Rayleigh-Taylor instabilities, showed that charged grains which are kinematically-decoupled from 
the ejecta gas might be impeded to traverse the contact discontinuity which separates the shocked ejecta from the shocked ISM. 

To offer further insights, it is crucial to tackle the evolution of dusty SNRs within young massive stellar clusters, \textit{where prolific 
dust enrichment is indeed localized} \citep{Consiglioetal2016,Leroyetal2018} and \textit{most massive stars are formed} \citep{LadaandLada2003}, 
with only a small percentage of them being ejected as runaways \citep{Khorramietal2016,Portegies-Zwartetal2010}. 

Nonetheless, additional complications arise. Firstly, the gas reinserted amidst these clusters is already shocked 
before the occurrence of first supernova explosion \citep[e.g.][]{Wunschetal2017} and thus the corresponding sputtering rate is high. 
Secondly, their high SN rate implies that the cluster and its surroundings will be frequently crossed by successive SN forward shocks 
(FSs), fostering the destruction of some fraction of the dust (from all sources) in the vicinity. Thirdly, the magnetohydrodynamic 
turbulence which is driven by the feedback of massive stars may trigger grain shattering \citep{Hirashitaetal2010}, producing an 
excess of small grains that can be rapidly sputtered in a thermalized medium. In this respect, 
\citet[][hereafter MST16 \& MWP17, respectively]{MartinezGonzalezetal2016,MartinezGonzalezetal2017} pointed out that the presence of 
near- to mid-infrared (NIR-MIR) excesses in SSCs, such as those observed in several clusters in M 33 and SBS 0335-052 \citep{Relanoetal2016,Reinesetal2008}, 
are likely to be an indication of dust efficiently produced by SNe, stochastically heated to high temperatures by electronic collisions and
photon absorptions, and readily destroyed in frequent ionic collisions. 

\citet[][]{TenorioTagleetal2015b} \citep[see also][]{Silichetal2017} have proposed a mechanism by which the SN yields 
(heavy metals) can be discharged out of a dense proto-globular cluster without polluting the gas within the cluster. Stressing that SN 
explosions are not synchronized events, they followed the evolution of SN blast waves in a strong density gradient. In the case of off-centered 
remnants, the interaction with the gas left-over from star formation will lead to the SNR's elongated growth and subsequently to a blowout phase: 
the development of Rayleigh-Taylor (RT) instabilities and shell fragmentation; thus allowing the venting of the SN ejecta to the ISM. 

The rapid post-blowout hydrodynamical evolution which leads to milder conditions (i.e. rapid decline in density and temperature) in SN ejecta, make 
the supernova blowout scenario a viable mechanism by which SN-condensed dust grains could be injected into the circumcluster medium before significant 
thermal sputtering takes place. As we are interested more on SSCs, here we will explore a similar scenario by means of three-dimensional hydrodynamical
simulations, which will enable us to follow the development and growth of RT instabilities and the breaking of the SNR's spherical symmetry. This scheme 
will also permit us to include the further dust processing which occurs when the SN reverse shock reaches the center of an SNR and generates a secondary
forward shock \citep{TenorioTagleetal1990,Micelottaetal2016}; while we will not, however, take into account the full dynamics of the grains relative to 
the gas. Furthermore, we will include the effects of the additional radiative cooling mediated by frequent collisions between gas particles and dust 
grains in such extreme conditions.

The main difference with the hydrodynamical model presented by \citet[][]{TenorioTagleetal2015b} is that the SNRs in the present work will 
interact with mass-loaded shocked stellar winds in an SSC rather than with the dense gas ($\sim 10^7$ cm$^{-3}$) in a still-embedded proto-globular 
cluster.

The paper is organized as follows: in Section \ref{sec:scheme} we formulate the star cluster model, the implementation of the
star cluster wind and the insertion of SNe with the help of three-dimensional hydrodynamical simulations; then we briefly describe the 
injection and destruction of dust grains in our simulations and the additional cooling induced by their presence. In Section 
\ref{sec:models}, we introduce several models to test the relevance of the rapid evolution of isolated SNRs and SNRs within SSCs
and the various implications regarding the efficiency of dust injection into the unshocked ISM. Finally, in Section \ref{sec:conclusions} we 
discuss our results and outline our main conclusions.

\section{Dusty Supernovae within a Star Cluster Wind Environment}
\label{sec:scheme}

\subsection{The Star Cluster Wind}
\label{sec:wind}

In a young massive stellar cluster (also called super star clusters, SSCs\footnote{SSCs 
are high-density coeval young stellar systems which stand out for their large stellar masses ($\sim 10^{4}$ to $10^{6-7}$ M$_{\odot}$) and compact radii of a 
few parsecs \citep{Whitmore2000}.}), the winds of individual massive stars 
collide and merge to establish a smooth star cluster wind \citep[see e.g.][]{ChevalierClegg1985,Silichetal2004,Silichetal2011,Palousetal2013}. 
Following \citetalias[][ and references therein]{MartinezGonzalezetal2016,MartinezGonzalezetal2017}, here we consider non-uniform coeval 
SSCs with a radial stellar density distribution of the form \mbox{$\rho_* \propto [1+(r/R_{c})^2]^{-\beta}$} truncated at a radius $R_{SC}$; 
where $r$ is the distance from the cluster center, $R_{c}$ is the core radius and $\beta$ is a constant. The total mass deposition rate via stellar 
winds within the cluster is $\dot{M} = 2 \eta_{he}L_{SC}/[(1+\eta_{ml}) V_{A\infty}^2]$, where $L_{SC}$ and $V_{A\infty}$ are the deposited mechanical luminosity and the adiabatic wind terminal speed, respectively. 
The heating efficiency ($\eta_{he}$), is the fraction of $L_{SC}$ converted into the wind 
thermal energy \citep[see e.g.][]{Silichetal2007} and the mass-loading parameter ($\eta_{ml}$), accounts for the amount of 
gas which was left-over by the cluster formation and that is incorporated into wind \citep[see e.g.][]{Wunschetal2011}. $L_{SC}$ is assumed 
to scale with the cluster's stellar mass, $M_{SC}$, as $L_{SC}= 3 \times 10^{40} (M_{SC}/10^6 \mbox{M}_\odot)$ erg s$^{-1}$ \citep{Leithereretal1999}.

We have carried out a series of three-dimensional hydrodynamical simulations of different star cluster winds using the adaptive mesh refinement code 
FLASH v4.3 \citep{Fryxelletal2000}. A modified version of the Piecewise Parabolic Method \citep[PPM,][]{ColellaandWoodward1984} is used to
solve the hydrodynamic equations. The numerical scheme includes the gravitational field of the star cluster, the self-gravity of the reinserted gas 
calculated by the tree-based solver described by \citet{Wunschetal2018} and the equilibrium cooling function for an optically thin plasma \citep{Schureetal2009}. All 
the simulations, unless otherwise stated, were carried out at high resolution (in a uniform $512^3$ grid) to ensure the correct computation of the gas radiative 
cooling, including that induced by dust grains, and the onset and growth of RT instabilities. The computational domain in our models is inscribed in a cube 
($60$\,pc)$^3$ in Cartesian geometry and the outer boundary conditions were all set to outflow. Further details on the numerical implementation of the cluster wind 
can be found in \citet{Wunschetal2017}.

\subsection{Supernovae}
\label{sec:sne}

The onset of the supernova era in coeval star clusters occurs some time around $\sim 3$ Myr (at the moment the most massive stars in the
cluster explode) and lasts for about $\sim 40$ Myr \citep[e.g.][]{Meynetetal1994}; at the time of each explosion a certain mass (gas and dust), 
$M_{ej}$, and kinetic energy, $E_{SN}$, are inserted into the simulation in a small spherical region with radius $R_{SN}$ 
and centered at an arbitrary position $(x,y,z)$. The initial ejecta gas mass density and velocity profiles are then assumed to be

\begin{eqnarray}
\label{eq:ejecta}
 \rho_{ej}=\frac{(3-\omega)}{4\pi}\frac{M_{ej}}{R_{SN}^3}\left(\frac{R_{SN}}{d} \right)^\omega ,\,\, 0 \leq \omega < 3   ,
\end{eqnarray}

and 

\begin{eqnarray}
\label{eq:vel}
 v_{ej} = \left( 2\frac{(5-\omega)}{(3-\omega)}\frac{E_{SN}}{M_{ej}} \right)^{1/2}\left(\frac{d}{R_{SN}} \right) ,
\end{eqnarray}

where $d$ is the distance measured from the center of the explosion and $\omega$ is a constant which we have chosen 
to be $0 \leq \omega < 3$ in order to have a finite enclosed mass without requiring to define a core, which is needed 
for $\omega \geq 3$ \citep[][see Appendix \ref{app:A2} for additional cases with steeper ejecta density profiles]{TangChevalier2017}. 
The ejecta is initially set to have a uniform temperature of $10^4$ K. We note that in the case of an SN occurring within 
the volume of an SSC, thermal sputtering is boosted due to the interaction of the SN ejecta with the shocked stellar winds 
of the massive stars that are eventually engulfed by the remnant. The presence of inhomogeneities in SN ejecta, which favor 
the formation and coagulation of dust grains \citep{BiscaroCherchneff2016,SarangiCherchneff2015}, is modeled as white noise, 
i.e. random initial density perturbations are generated to mimic the presence of a collection of clumps in the remnant.

\subsection{Dust Injection}
\label{sub:injection}

\begin{figure*}
\includegraphics[width=\linewidth]{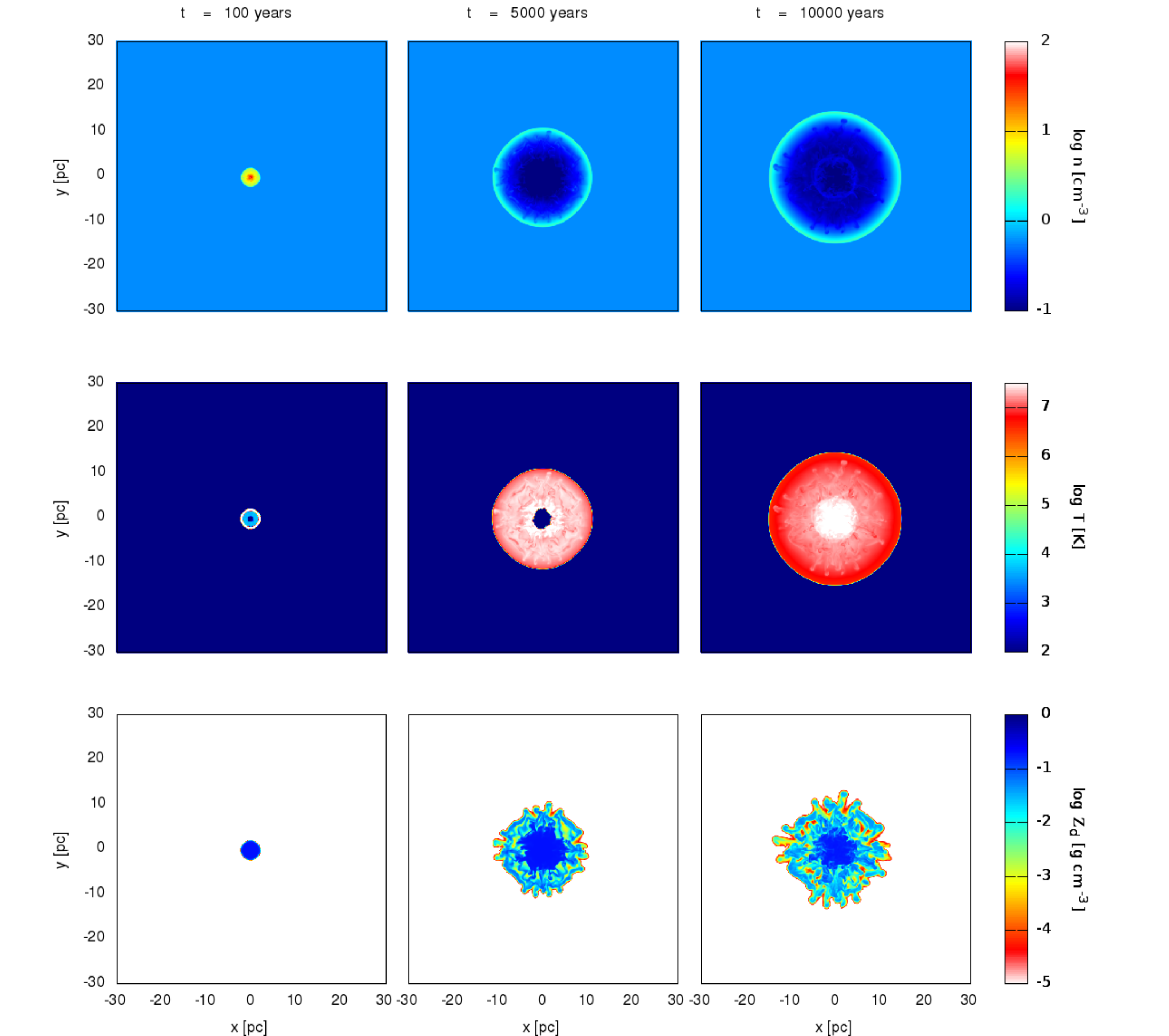}
\caption{The evolution of an SNR in the diffuse medium. The upper, middle and bottom panels 
display slides in the plane $z=0$ of the distribution of gas number density, gas temperature, and dust-to-gas mass ratio 
at three different post-explosion times (left: 100 years, center: 5000 years, right: 10000 years). In this case, the assumed initial ejecta 
density and velocity profiles corresponds to $\omega=1$.}
\label{fig:1}
\end{figure*}

We have developed CINDER (Cooling INduced by Dust \& Erosion Rates); a module for three-dimensional hydrodynamical 
simulations in FLASH which on-the-fly computes the rates of radiative cooling and thermal sputtering due to gas-grain collisions in a given grid cell 
provided an initial distribution of grain sizes and dust content. With CINDER we can follow the evolution of the dust mass injected by SNe and subjected 
to shock-processing and, to the best of our knowledge, it provides the first implementation of the cooling induced by dust in a hot plasma suitable for 
time-dependent three-dimensional hydrodynamical simulations. For the goals of the present work, dust is instantaneously formed at the time of the respective 
SN explosion and traced via the local dust-to-gas mass ratio as an advecting mass scalar associated to a representative grain size (in other words, one 
advecting field array per representative grain size). The grain sizes are initially distributed according to a power-law of the form $\sim a^{-\alpha}$, 
where $a$ is the grain radius, with lower and upper limits $a_{min}=1$ nm and $a_{max}=0.5$ $\mu$m, respectively, and $\alpha=3.5$ 
\citep[][hereafter MRN]{MRN1977}. We have logarithmically-binned the grain size distribution; thus the representative value of the $i$th grain size bin, is defined as

\begin{eqnarray}
 a_{m}^{(i)} = \sqrt{ a_{min}^{(i)} \displaystyle a_{max}^{(i)} }  ,\,\,  1 \leq i \leq N_{bins}   ,
\end{eqnarray}

where $a_{min}^{(i)}$ and $a_{max}^{(i)}$ are the lower and upper limits of the respective grain size bin which are given by

\begin{eqnarray}
 a_{min}^{(i)} = a_{min} \times 10^{(i-1) \delta a / N_{bins}},
\end{eqnarray}

and 

\begin{eqnarray}
 a_{max}^{(i)} = a_{min} \times 10^{i \delta a / N_{bins}}.
\end{eqnarray}

In the above equations, $N_{bins}$ is the total number of grain size bins and $\delta a$ is equal to $\log{a_{max}}-\log{a_{min}}$. Note that 
the representative grain sizes are equidistant in the logarithmic scale and that the smaller $i$ is, the smaller the represented grain size. 
Hence, the initial dust mass fraction corresponding to the $i$th grain size bin is

\begin{eqnarray}
\label{eq:frac}
 f_{M_{d}}^{(i)} = \displaystyle \frac{(a_{max}^{(i)})^{-\alpha+4}-(a_{min}^{(i)})^{-\alpha+4}}{a_{max}^{-\alpha+4}-a_{min}^{-\alpha+4}} ,
\end{eqnarray}

which shows that the dust mass is dominated by large grains for $\alpha<4$. The total dust mass, $M_{d}$, for all grain size bins has 
been halved into silicate and graphite grains, with grain densities equal to $\rho_{gr}=3.3$ g cm$^{-3}$ and 
$\rho_{gr}=2.26$ g cm$^{-3}$, respectively. 

Grains immersed into a plasma at temperatures $\gtrsim 10^6$ experience thermal sputtering due to frequent ionic 
collisions which remove a certain amount of mass from their surfaces. Since the mass of individual grains, $m_{gr}$, 
decreases as $\dot{m}_{gr} = 3 m_{gr}\dot{a}/a$, where $\dot{a}$ is the rate of decrease in the grain size, after a small 
time, $\Delta t$ (a few years in our simulations), the local dust-to-gas mass ratio of the respective size bin, $Z_{d}^{(i)}$, is 
reduced according to \citep{McKinnonetal2017}

\begin{eqnarray}
 Z_{d}^{(i)}(t+\Delta t) &=& Z_{d}^{(i)}(t) 
  \left( 1 - \displaystyle \frac{3|\dot{a}|\Delta t}{a_{m}^{(i)}} \right) ,
\end{eqnarray}

where $\dot{a}$ is given by \citep{TsaiMathews1995}

\begin{eqnarray}
      \label{eq:A3}
\dot{a} = -1.4 n h \left[\left(\frac{10^6 \mbox{K}}{T}\right)^{2.5}+1 \right]^{-1}   ,
\end{eqnarray}

and depends on the gas number density, $n$, and the gas temperature, $T$. Additionally, $h$ is a constant equal to $3.2\times 10^{-18}$ cm$^{4}$ s$^{-1}$. 
The above relation approximates the detailed computation of $\dot{a}$ obtained by \citet{DraineandSalpeter1979} and \citet{Tielensetal1994} for both graphite and 
silicate grains. As in \citetalias{MartinezGonzalezetal2017}, we have corrected equation \eqref{eq:A3} to reflect the size-dependence of the 
sputtering yield as prescribed by \citet{SerraDiazCanoetal2008} and \citet{Bocchioetal2012}. If it occurs that
$a_{m}^{(i)}-|\dot{a}|\Delta t<a_{min}^{(i)}$, the grains in the bin are not well represented by $a_{m}^{(i)}$ anymore, 
and the mass of the $i$th size bin is transfered to the lower adjacent bin. Only in the case of $i=1$, and provided the 
above condition is fulfilled, the mass in the bin is considered to be fully returned to the gas phase. 

The additional gas cooling provided by the presence of dust grains via electronic and ionic collisions is 
calculated following the formulation given by \citet[][see also Appendix A.1 in \citealt{MartinezGonzalezetal2016}]{Dwek1987}
by summing up the contributions from individual grain size bins. We have checked the value of $N_{bins}$ necessary to achieve 
a good agreement between the semi-analytic calculation of the cooling rate \citepalias{MartinezGonzalezetal2016} and the discrete 
calculation after logarithmically-binning the grain size distribution. With $N_{bins}=6$ the agreement is already excellent; 
however, as thermal sputtering can completely erode small grains much faster than the large ones, we have set $N_{bins}=10$ to 
adequately sample the population of small grains and their associated sputtering rates. The relevance of dust-induced radiative cooling 
is that, while ionic collisions are very efficient at destroying the small grains, these grains are also efficiently heated by electronic 
collisions, stochastically fluctuate in temperature, and manifest themselves mainly at NIR-MIR wavelengths 
\citepalias{MartinezGonzalezetal2016,MartinezGonzalezetal2017}; small grains may have short lifetimes but their imprint on the gas 
thermodynamics and the survival of larger grains remains for a longer time.

Additionally, the metallicity of the reinserted gas (both winds and SN matter) has been chosen to be solar in order to keep consistency 
with the value of $\dot{a}$ in equation \ref{eq:A3}, which is only valid for a gas with solar composition. Nonetheless, as pointed out 
by \citet{Nathetal2008}, the metallicity of SN ejecta exceeds the solar value, and the sputtering yields will be higher as
a result of more frequent collisions with heavy ions. 

Among the limitations of our modeling is the arbitrary choice of equal mass fractions of silicate and graphite grains 
\citep[which has been used to fit the ejecta mass of Cas A, e.g.][]{DeLoozeetal2017,Bevanetal2016}. Carbonaceous grains might be 
more efficiently destroyed than silicates \citep{Jonesetal2011}; however, as we have used identical thermal sputtering rates, size 
distribution and dynamics for both species, their mass fractions would not change during our simulations. Following the evolution of the 
grain species separately would increase computational costs and the complexity of the problem and, for now, it is beyond the scope of the paper. 

Another limitation is the use of the classical \citealt{MRN1977} grain size distribution which was originally derived for already-processed ISM dust. 
Different types of core-collapse SNe tend to form grains of different characteristic sizes; for instance, type II-B SNe, with thin hydrogen 
envelopes and high expansion velocities, form grains weighted more towards small sizes ($\lesssim 0.006 \mu$m), while Type II-P SNe, with massive 
hydrogen envelopes and low expansion velocities, tend to condense large grains ($\sim 0.1 \mu$m) which can survive longer 
\citep{Kozasaetal2009, Nozawaetal2010}. For this reason, in Appendix \ref{app:A1} we have also considered the case of a log-normal grain size 
distribution dominated by large grains in which the surviving dust mass fraction increases.

\section{Models}
\label{sec:models}
We have defined several models in order to quantify the fraction of SN-condensed dust that is injected into the ISM after 
shock-processing. In \ref{sec:isolated}, we first investigate two cases of SN explosions occurring in the diffuse ISM with a constant 
number density. These two cases, our fiducial runs, differ only in their initial ejecta density distribution (homogeneous vs. 
stratified). Then, in Section \ref{sec:SNRwind} we explore the interaction of different SNRs with the shocked winds of neighboring 
massive stars in a star cluster with mass $10^5$ M$_{\odot}$. In these cases we do not follow the occurrence of successive SN 
explosions, as it is done in Section \ref{sec:SNR-SNR}, where a more massive cluster ($10^6$ M$_{\odot}$) is considered.

\subsection{An SNR evolving in the diffuse ISM}
\label{sec:isolated}

\begin{figure}
\includegraphics[width=\linewidth]{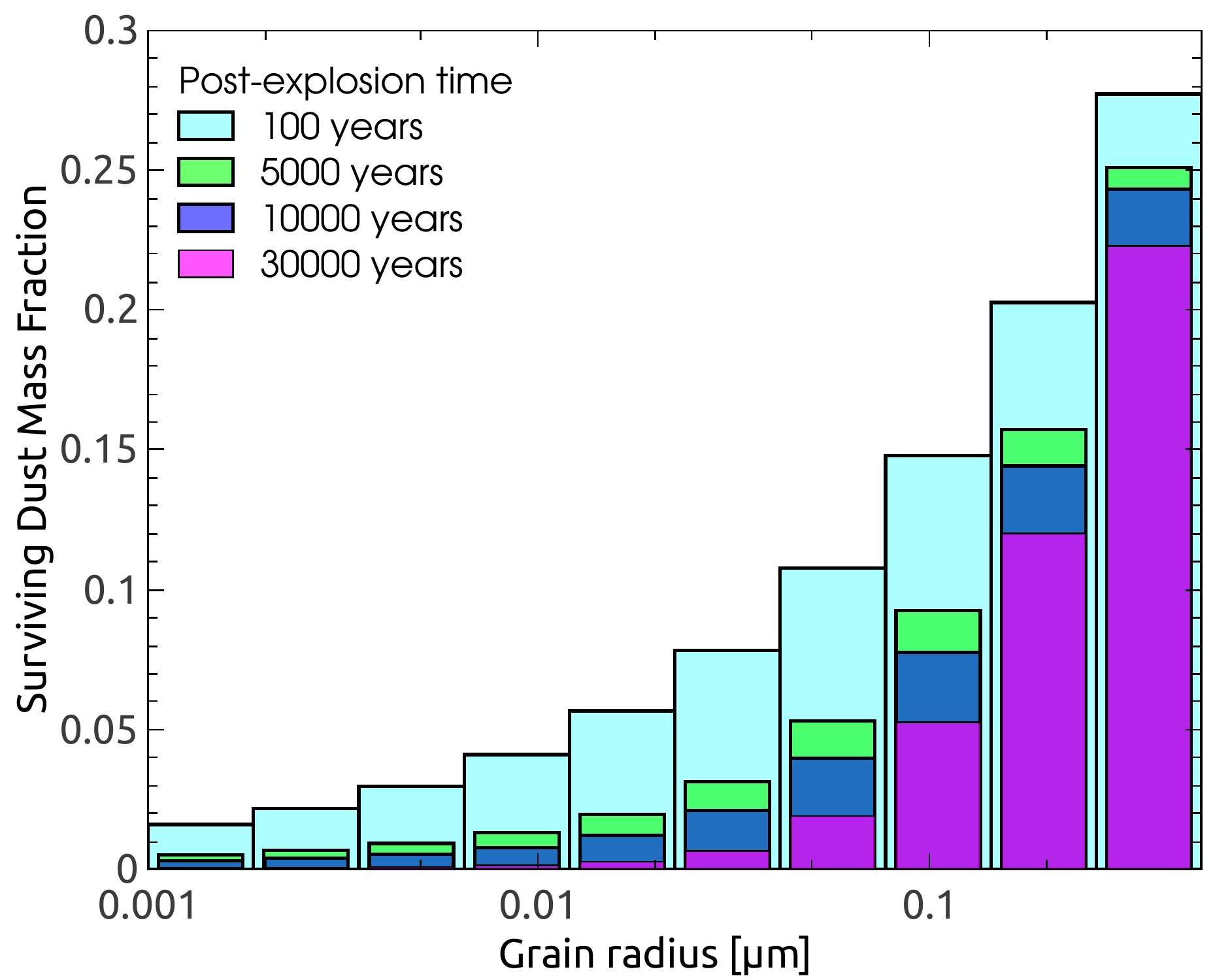}
\caption{Dust mass evolution per grain size bin in our stratified fiducial run ($\omega=1$). The histograms show the distribution of dust mass 
(summed over all the grid cells including both graphites and silicates) per size bin at different post-explosion times (light blue: 100 years, 
green: 5000 years, dark blue: 10000 years, violet: 30000 years).}
\label{fig:2}
\end{figure}

For our fiducial runs, we consider a medium of constant number density equal to $1$ cm$^{-3}$. The SN ejecta is initially distributed 
according to a homogeneous and (a particular case of) a stratified ejecta density profile ($\omega=0$ and $\omega=1$, respectively, see 
equations \eqref{eq:ejecta} and \eqref{eq:vel}). A stratified density profile can be regarded as more realistic given the inner structure of 
the progenitor star prior to explosion and how shock acceleration proceeds during the explosion \citep{ChevalierandFransson1994}. 
Nevertheless, the homogeneous case allows a better comparison with the results obtained by \citetalias{Bocchioetal2016} as it is the case they 
explored. 

\begin{figure}
\includegraphics[width=\linewidth]{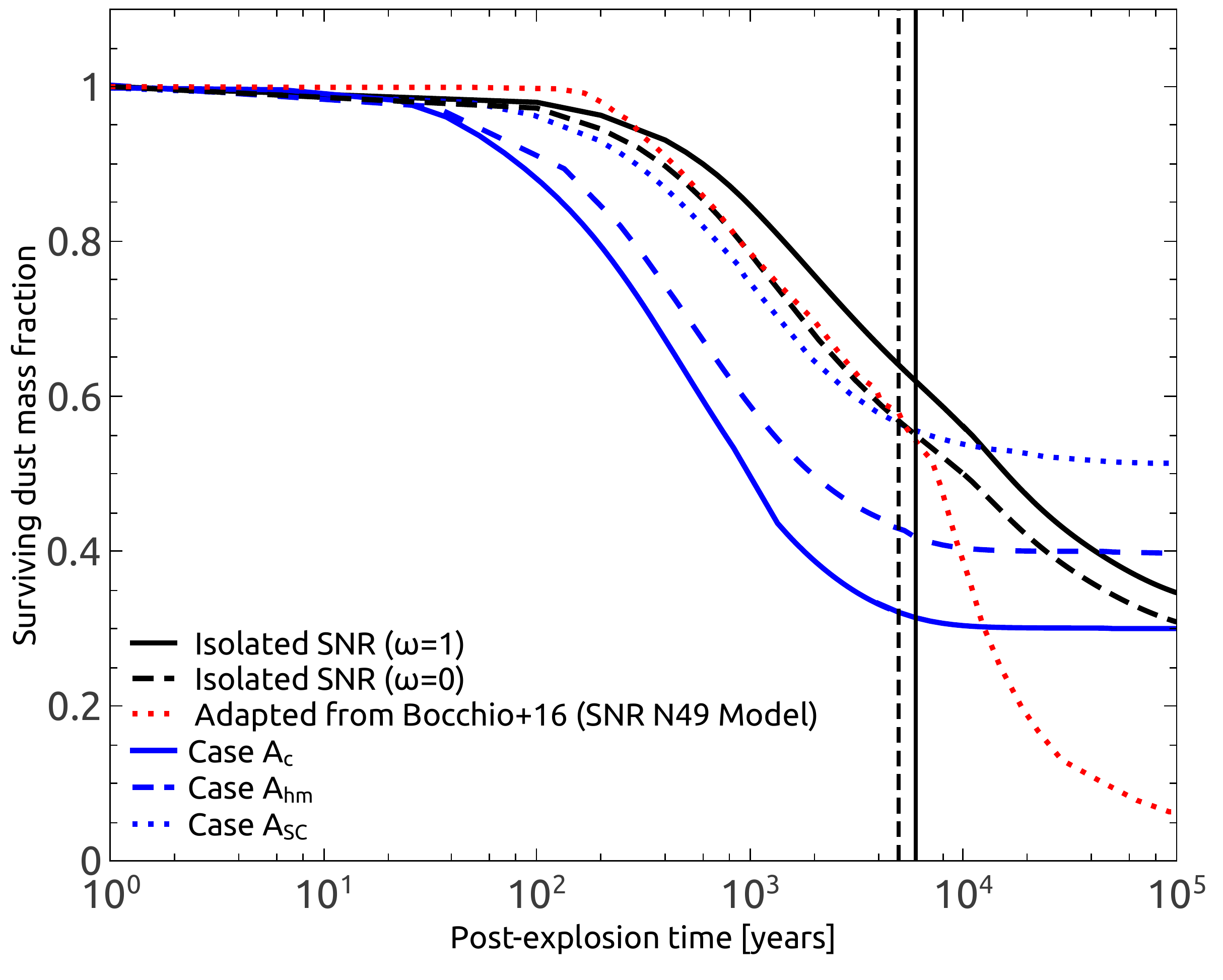}
\caption{Evolution of the dust mass after its insertion in single (model \textit{A}) SN explosions. The blue curves correspond to model \textit{A} 
(cases \textit{A$_{c}$} (solid line), \textit{A$_{hm}$} (dashed line) and \textit{A$_{SC}$} (dotted line)). For reference, the solid ($\omega=1$) 
and dashed ($\omega=0$) lines represent the evolution of the dust mass in the case of an SNR evolving in the diffuse ISM with constant density, 
while the dash-dotted red line was adapted from \citet{Bocchioetal2016} model for the SNR N49. Vertical lines mark the times when the RS reaches the 
center of the SN ejecta in the isolated cases.}
\label{fig:3}
\end{figure}

The ejecta mass in all cases is composed of $1.0$ M$_{\odot}$ of dust and $5$ M$_{\odot}$ of gas, $M_{ej}=M_{d}+M_{gas}$, and the kinetic energy 
of the explosion is taken to be $E_{SN}=10^{51}$ erg. In Figure \ref{fig:1} we show slides in the plane $z=0$ of the distribution of gas number density, gas temperature, and 
dust-to-gas mass ratio at post-explosion times $0.1$, $5$ and $10$ kyr. In Figure \ref{fig:2} we present histograms of the total mass of dust in the ejecta per grain size bin at the same post explosion 
times as in Figure \ref{fig:1}, and at $30$ kyr. While the shape of the FS remains spherical at all times, RT instabilities develop in the decelerating SN 
ejecta, thus causing it to depart from spherical symmetry. The remnant enters the Sedov-Taylor phase after the swept-up mass surpasses by a factor of several tens the 
ejected mass \citep{Gull1973} and subsequently, the SN ejecta becomes fully thermalized at $\sim 6$ kyr post-explosion. When reaching the center of the SNR, 
the reverse shock bounces and creates a secondary forward shock (sFS) which eventually catches up with the leading forward shock \citep{TenorioTagleetal1990}, 
in our case at $\sim 32$ kyr. The sFs can be identified in the snapshots at $10$ kyr in Figure \ref{fig:1} as the innermost shock-heated region 
with radius $\sim 5$ pc. We predict a somewhat larger percentage of surviving dust mass than those of \citetalias{Bocchioetal2016} ($\sim 1-8\%$ 
versus $\sim 25-30\%$ in our fiducial runs). This difference results from taking into account the dynamics of the grains relative to 
the gas, a different size distribution and chemical composition of the grains. Nevertheless, as observed in Figure \ref{fig:3}, the predicted 
slope in the evolution of the dust mass is remarkably similar in both approaches (see the solid ($\omega=1$) and dashed ($\omega=0$) black 
lines which display our fiducial runs versus the 'SNR N49' model results obtained by \citetalias{Bocchioetal2016} (red dotted line)) between 
$\sim 0.6$ and $10$ kyr. At times $\gtrsim 10$ kyr, they found that a significant fraction of the kinematically-decoupled grains 
reside between the contact discontinuity and the FS, where the efficiency of thermal sputtering is the highest. As in our models the grains
are perfectly coupled to the ejecta and therefore do not traverse the contact discontinuity, we are unable to reproduce the sudden fall in the 
dust mass obtained by \citetalias{Bocchioetal2016} at times $\gtrsim 10$ kyr.

However, \citet{Fryetal2018}, using 1-D+ hydrodynamical simulations, found that turbulent ISM magnetic fields, mostly amplified between the 
contact discontinuity and the FS, might prevent kinematically-decoupled charged grains to traverse the contact discontinuity to instead being 
reflected within the ejecta. As a result, the sudden increase in dust destruction in \citetalias{Bocchioetal2016}'s calculations at times 
$\gtrsim 10$ kyr may not occur given that the grains would never experience the harsher conditions ahead of the contact discontinuity. Moreover, 
\citet{Fryetal2018} also found that drag and kinetic sputtering are of relatively minor relevance since the velocity of large ($\sim 0.1$ $\mu$m) 
grains with respect to the gas is $\sim 175$ km s$^{-1}$ \citep[see Figure 2 in ][for a comparison of the thermal and 
kinetic sputtering rates as a function of gas temperature and gas-grain relative velocity]{Goodsonetal2016}. They also stressed the importance 
of RT instabilities in the survival of SN-condensed grains as they influence the location of the RS. In summary, even though we have not included all the 
physical processes that \citet{Bocchioetal2016} and \citet{Fryetal2018} have already considered, the main advantages of our three-dimensional 
approach are that RT instabilities appear naturally; that we have considered radiative cooling, including that induced by gas-grain 
collisions; and the enhancement in thermal sputtering mediated by the sFS. 

\subsection{SNRs interacting with the shocked winds of massive stars}
\label{sec:SNRwind}

\begin{figure}
\includegraphics[width=\linewidth]{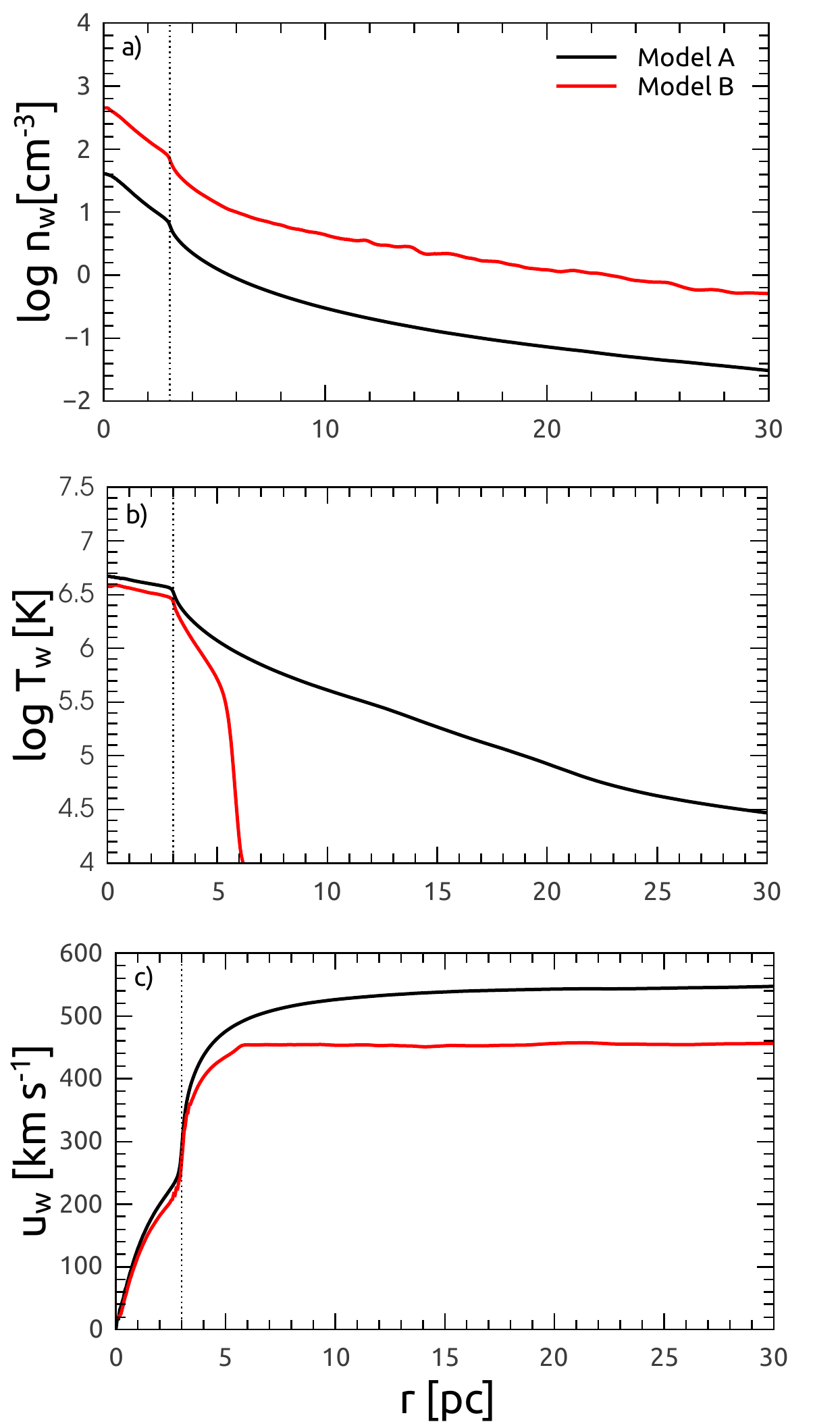}
\caption{The wind radial profiles before SNe. The top (a), middle (b) and bottom (c) panels show the radial distribution of the wind number density, 
temperature and velocity for our star cluster wind models \textit{A} (black lines) and \textit{B} (red lines), respectively. The vertical 
dotted-lines mark the star cluster cut-off radius, $R_{SC}$.}
\label{fig:4}
\end{figure}

\begin{figure*}
\includegraphics[width=\linewidth]{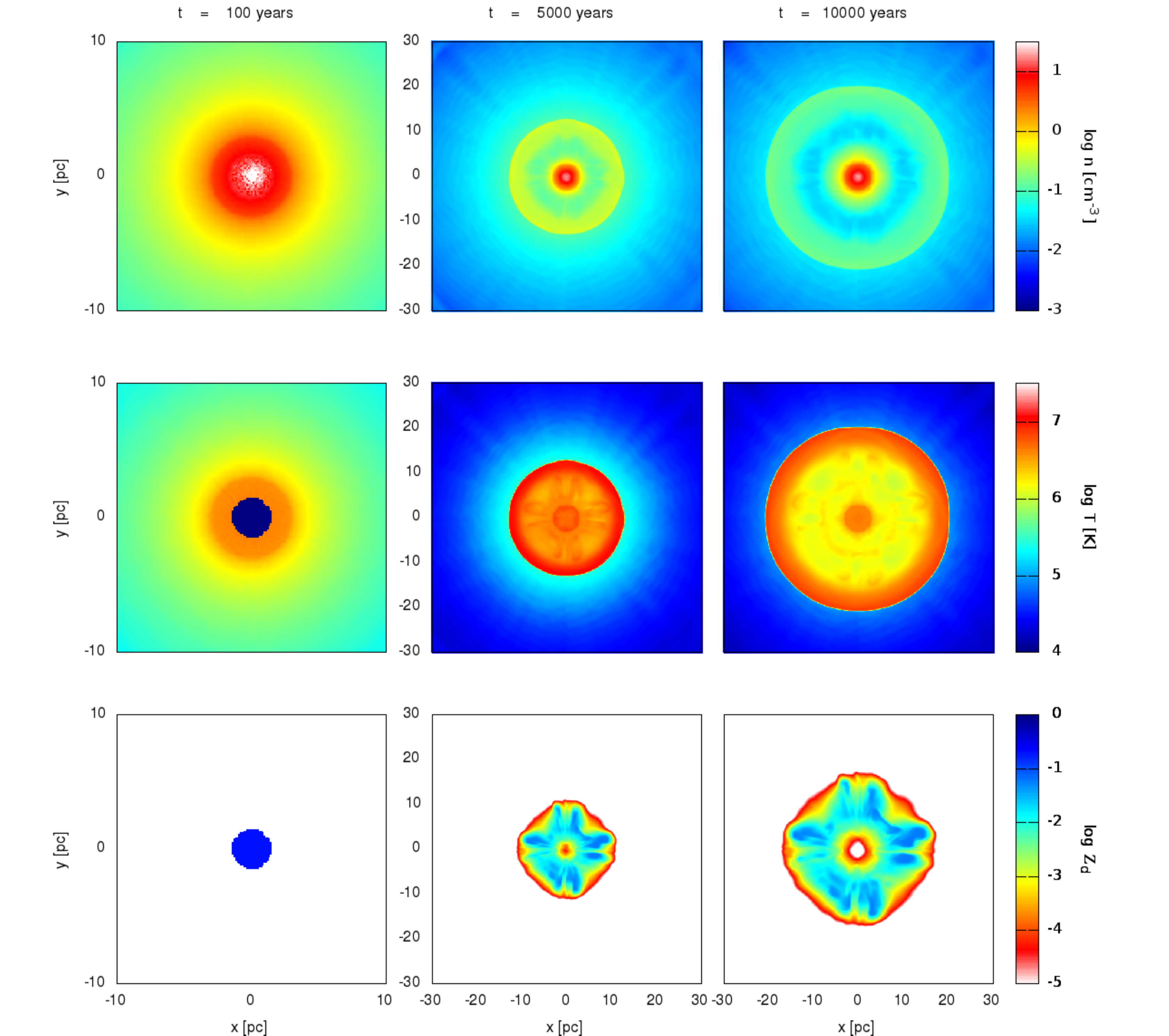}
\caption{The evolution of an SNR exploding at the center of an SSC with mass $10^5$M$_{\odot}$ in case \textit{A$_{c}$}. The upper, middle and 
bottom panels display slides in the plane $z=0$ of the distribution of gas number density, gas temperature, and dust mass density at three different 
post-explosion times (left: 100 years, center: 5000 years, right: 10000 years). Note the different length scale in the left column with 
respect to the central and right columns.}
\label{fig:5}
\end{figure*}

\begin{figure*}
\includegraphics[width=\linewidth]{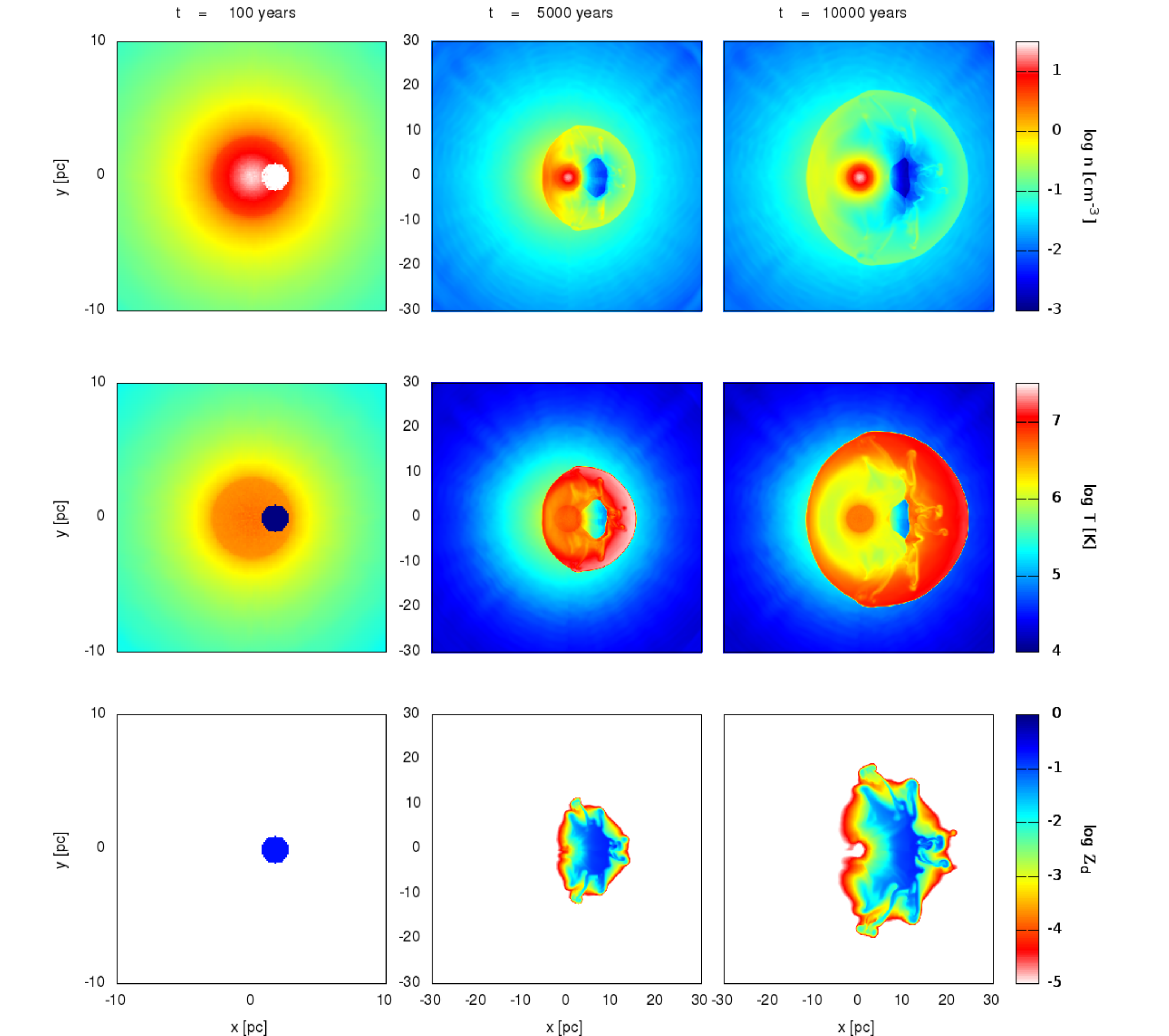}
\caption{Same as in Figure \ref{fig:5} but for case \textit{A$_{hm}$}.}
\label{fig:6}
\end{figure*}

\begin{figure*}
\includegraphics[width=\linewidth]{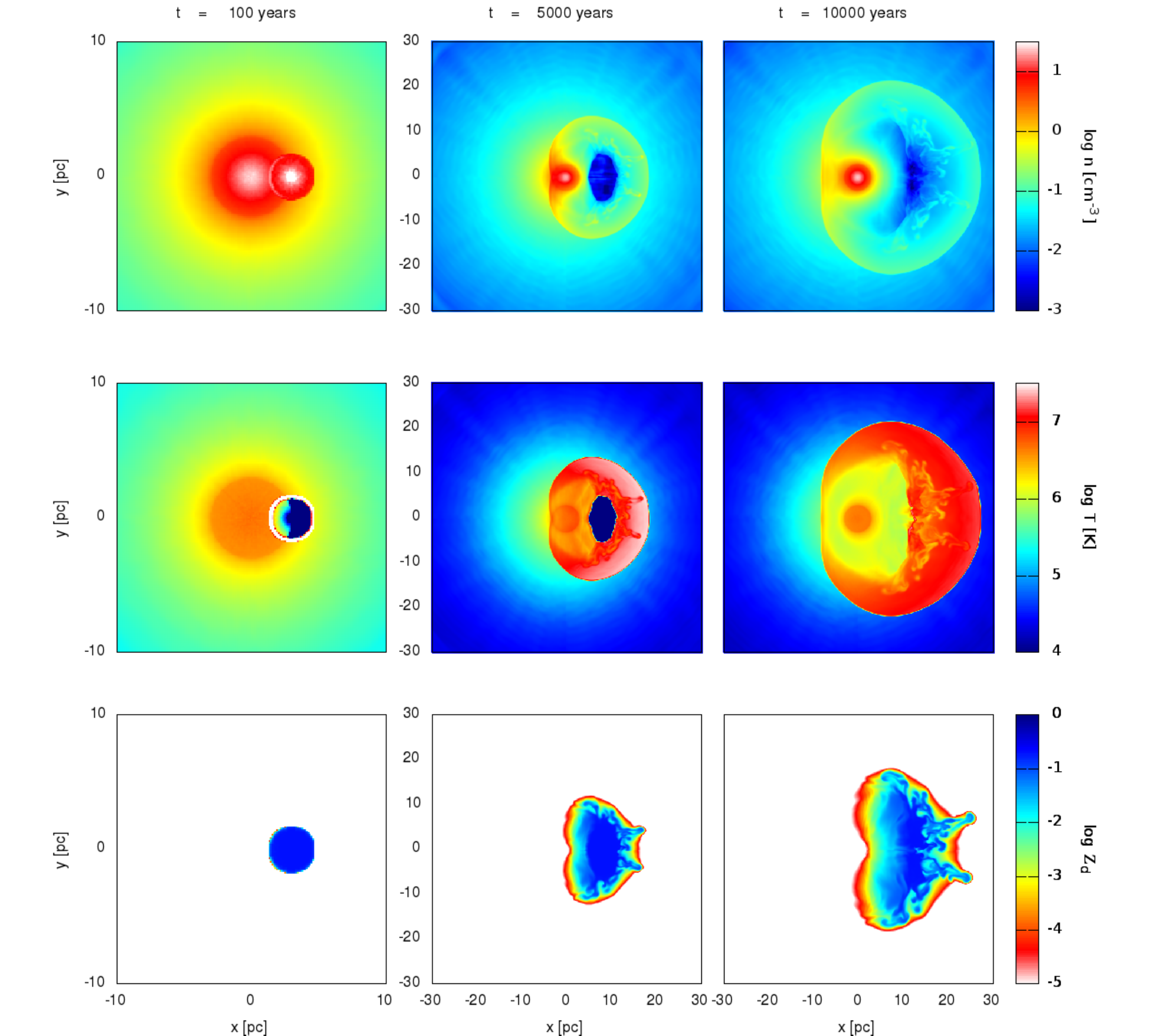}
\caption{Same as in Figure \ref{fig:5} but for case \textit{A$_{SC}$}.}
\label{fig:7}
\end{figure*}

Our attention is now focused on the evolution of dusty SNRs within young massive stellar clusters. From the large number of parameters, here we 
will explore the impact of varying the star cluster mass and position of the SN explosion within the cluster. In all models we have fixed the 
physical dimensions and density profile of the clusters ($R_{c}=1$ pc, $R_{SC}=3$ pc and $\beta=3/2$, which imply that the half-mass radius 
is $R_{hm}\approx 1.7$ pc), the adiabatic wind terminal speed ($V_{A\infty}=2500$ km s$^{-1}$), and the values of the heating 
efficiency \citep[$\eta_{he}=0.1$, which is of the order of that indicated by the small observed sizes and expansion velocities 
of H\,{\sc ii} regions associated to SSCs in M 82][]{Silichetal2007,Silichetal2009} and mass-loading \citep[$\eta_{ml}=1.0$,][]{Wunschetal2017}. 
The effect on the hydrodynamical wind solutions provided by varying these parameters has been discussed by us elsewhere 
\citep[e.g.][]{Silichetal2011,Palousetal2013,MartinezGonzalezetal2016,TenorioTagleetal2013,Wunschetal2011,Wunschetal2017}. 
To briefly summarize the relevance of the additional input parameters, thermal sputtering will be boosted (or decreased) in the case of a denser
(lighter) cluster wind, which will occur in the case of a more compact (extended) star cluster enclosing the same stellar mass, and/or smaller 
(larger) values of $V_{A\infty}$ and $\eta_{he}$, and/or larger (smaller) values of $\eta_{ml}$.

Our first star cluster model, hereafter model \textit{A}, considers a total stellar mass equal to $10^5$ M$_{\odot}$ 
($L_{SC}=3 \times 10^{39}$ erg s$^{-1}$) (see Figure \ref{fig:4} which shows the averaged radial distribution of 
wind number density, temperature and velocity for the cluster wind calculated in model $A$ and in model $B$ defined in the 
following section). Three different cases regarding the position of the SN explosion are discussed 
(all assumed to occur in the $x$ axis of the Cartesian plane). In case \textit{A$_{SC}$}, the explosion occurs at the edge of the 
cluster; in case \textit{A$_{hm}$}, it occurs at a distance equal to the half-mass radius; and in case \textit{A$_{c}$}, the 
explosion occurs at the very center of the star cluster. The values of the mass (gas and dust) and kinetic energy deposited by each 
SN are the same as in our more realistic stratified fiducial run.

The distribution of the gas number density, temperature and dust-to-gas mass ratio at post-explosion times $0.1$, $5$ and $10$ kyr, 
respectively, for the three cases in model \textit{A} are shown in Figures \ref{fig:5} to \ref{fig:7}. In this scenario, especially in 
case \textit{A$_{hm}$}, the SN ejecta is rapidly thermalized when it interacts with the dense shocked matter reinserted by stellar winds 
within the cluster. As a result, the corresponding sputtering rate is high and dust destruction proceeds much faster than in the isolated 
SNR considered in the fiducial runs. However, the density gradient (particularly in the off-centered cases \textit{A$_{hm}$} and 
\textit{A$_{SC}$}) and the additional thermal energy injected by stellar winds \citep{Silichetal2017} facilitate the expansion 
of the remnant. The RS takes longer to reach the entire ejecta ($\sim 13$ and $10$ kyr, respectively), the conditions in the ejecta become 
milder and thermal sputtering ceases to be efficient at earlier times than in the fiducial runs. The combination of all these distinct 
physical processes allows a substantial increase in the surviving dust mass.

Since the size of the SNRs exceeds the computational domain after a few times $10^4$ years and it would be too computationally expensive
to follow the corresponding SNR evolution for a longer time (e.g. $\gtrsim 10^5$ years with a time-step of the order of years) in an extended 
computational domain at the highest resolution, we tested the dust mass evolution with a lower resolution ($256^3$) in a computational box of 
size $100$ pc. By doing so, the associated dust mass is $\lesssim 3\%$ higher at all times than in the $512^3$ runs, which gives us 
confidence to extrapolate the evolution of the dust mass with the expected behavior. Nevertheless, almost all dust destruction in the three 
cases occurs before the first $10^4$ years post-explosion. In case \textit{A$_{c}$} $\sim 30\%$ of the dust survives shock-processing 
(see the solid blue line in Figure \ref{fig:3}). In cases \textit{A$_{hm}$} and \textit{A$_{SC}$}, the remnants elongate, become RT unstable 
and the remaining dust mass asymptotically approaches $\sim 40\%$ and $\sim 50\%$, respectively (see the blue dashed and dotted lines in Figure 
\ref{fig:3}).

\subsection{Colliding Supernova Remnants}
\label{sec:SNR-SNR}

\begin{figure}
\epsscale{1.25}
\plotone{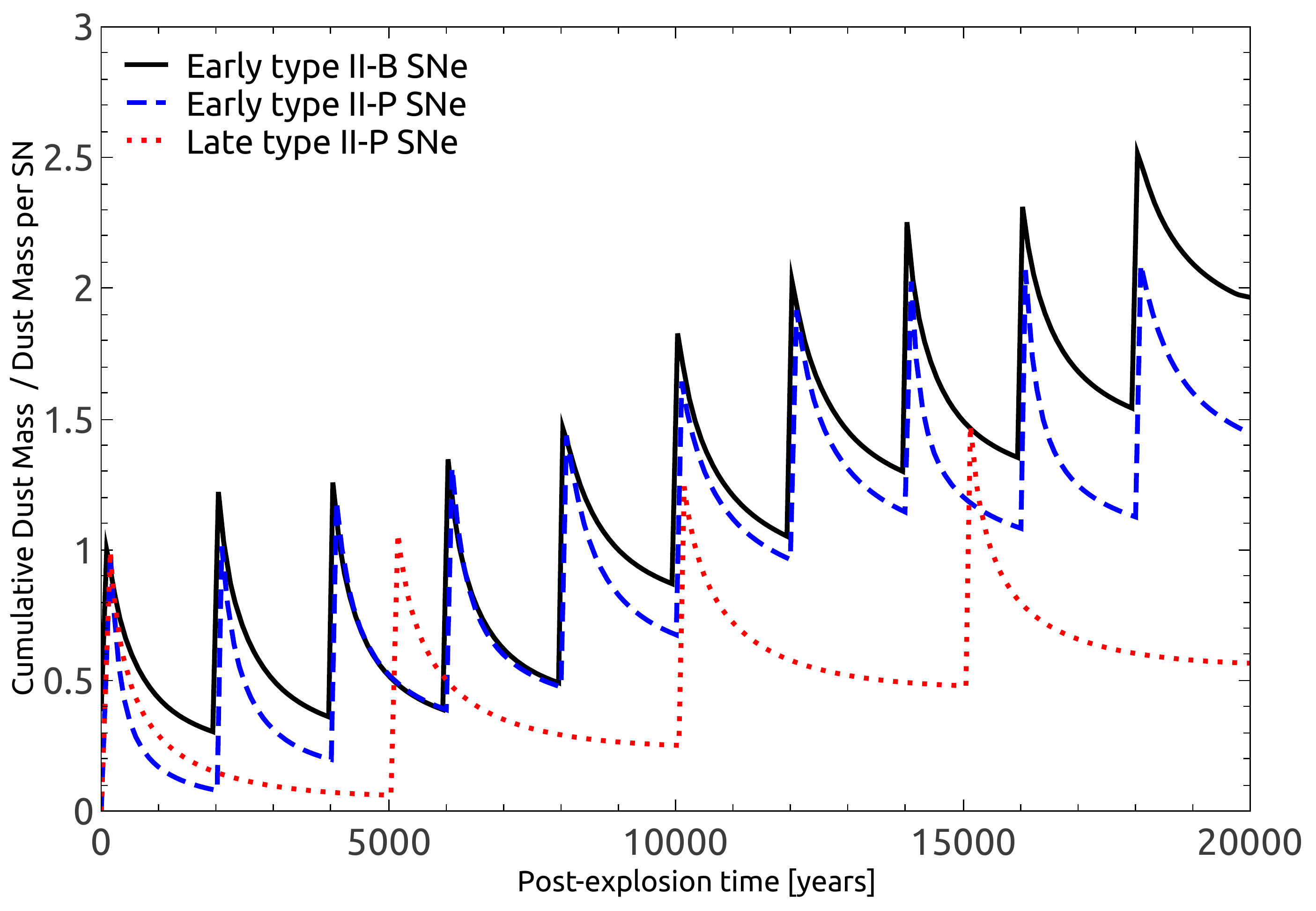}
\caption{Evolution of the cumulative dust mass (normalized to the average dust injected by SNe) after several sequential SN explosions 
(model \textit{B}) in our early type II-B (black solid line) and II-P SNe (blue dashed line) cases and late type II-P SNe (red dotted line) 
case. Note the tendency to increase the amount of dust even after multiple shock processing.}
\label{fig:8}
\end{figure}

\begin{figure*}
\includegraphics[width=\linewidth]{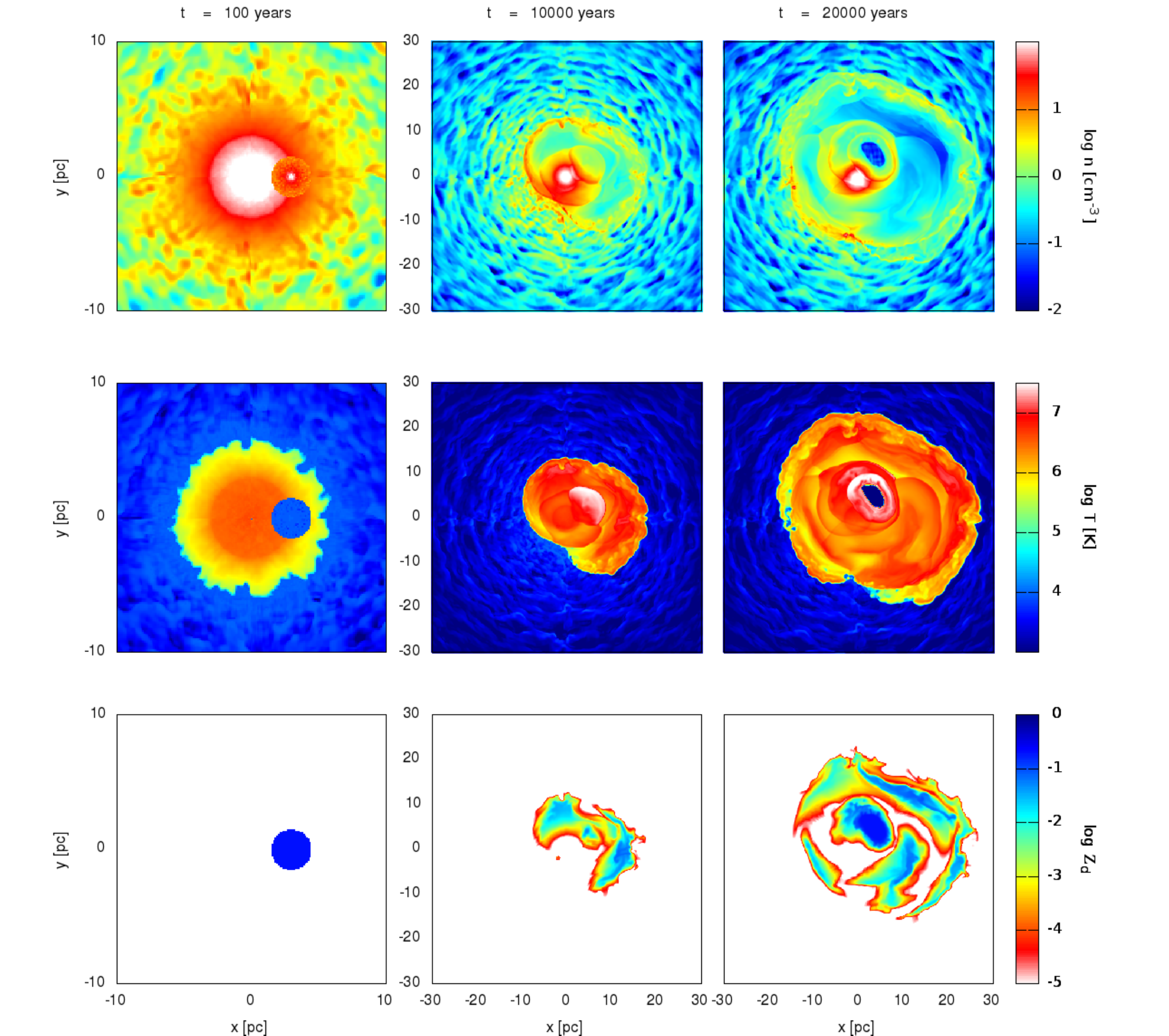}
\caption{Same as in Figure \ref{fig:5} but for the injection of dust by sequential SNe (model \textit{B}) in the early type II-B SNe case
at times 100 (left), 10000 (center), and 20000 years (right).}
\label{fig:9}
\end{figure*}

A larger stellar mass translates into a higher SN rate; if large/high enough, thermal sputtering may not only be triggered by the 
interaction with the shocked stellar winds and the passage of the RS/sFS, but also because of the interaction with subsequent SN 
forward shocks that catch up with previously ejected SN matter. Our model \textit{B} thus corresponds to a cluster with total stellar 
mass $10^6$ M$_{\odot}$, with SNe going off at random positions within the cluster. We first explore the case when the multiple 
SN explosions have the same input parameters as in the stratified fiducial run, each separated by $\sim2$ kyr according to the
respective SN rate which is $\sim10^{-3.3}$ yr$^{-1}$  during the early SN era (between $3.4$ and $5$ Myr of the age of the cluster for a 
\citetalias{Kroupa2001} initial mass function and Geneva evolutionary tracks \citep{Meynetetal1994}). This case, with a low value of the ejecta mass (5 M$_{\odot}$ 
of gas and 1 M$_{\odot}$ of dust) is appropriate for type II-B SNe like Cas A, with $\sim 4$ M$_{\odot}$ of gas \citep{HwangandLaming2012} 
and $\sim 1.1$ M$_{\odot}$ of dust \citep{Bevanetal2016}. We also study cases appropriate for type II-P SNe at two extremes: early SNe era with
the same SN rate as in the type II-B case, and late SN era ($\sim 40$ Myr) with one SN occurring every $\sim 5$ kyr. In the first 
extreme, only the most massive stars in the cluster explode, while in the second extreme the exploding stars correspond to $\sim 9$ M$_{\odot}$ 
progenitors. The values of the ejecta mass and kinetic energy for the type II-P explosions were taken from the calculations of 
\citet{Sukhboldetal2016}. In \citet{Sukhboldetal2016}, progenitors above $\sim 28$ M$_{\odot}$ hardly explode as SNe and even if they do, 
they are rare events. Therefore, for the early type II-P SNe, we take an average $27$ M$_{\odot}$ progenitor with ejecta mass $13.73$ M$_{\odot}$ 
and kinetic energy $1.08\times10^{51}$ erg; whereas for the late type II-P SNe, we consider an average $9$ M$_{\odot}$ progenitor with ejecta 
mass $7.38$ M$_{\odot}$ and kinetic energy $1.1\times 10^{50}$ erg.

In Figure \ref{fig:8} (and Figure \ref{fig:9} only for the early type II-B SNe case), we present $20$ kyr of the evolution of the dust mass 
injected by several sequential SNe within the star cluster in model \textit{B}. The trend observed in the Figures is that replenishment of dust 
by sequential SNe takes over grain destruction induced by the interaction with the shocked stellar winds, the reverse and secondary forward 
shocks and the multiple forward shocks originating from secondary SN explosions. In the type II-P cases, a denser and slower SNR expansion 
leads to an increase in the efficiency of dust destruction. The efficiency of destruction is also a strong function of the 
location of the secondary SN explosions, with antipodal explosions having the least effect on each other and therefore not all the subsequent 
forward shocks would cause a major surge in the destruction of dust. 

Even at the high SN rates explored, we predict the survival of $\sim 15-20\%$ of the total dust injected by SNe after multiple shock processing. 
The results presented here generalize those previously obtained by us \citep[][see also Figure 1 in \citealt{MartinezGonzalezetal2016}]{TenorioTagleetal2013} 
by means of 1-D semi-analytic hydrodynamical modeling where all SNe, interacting only with the shocked stellar winds, were assumed to occur at 
the cluster's center.

\subsection{Other physical processes}
\label{sec:decoupling}

We have not provided a self-consistent picture that includes dust formation, nor have we included the magnetohydrodynamic 
turbulence set by massive stars \citep[e.g.][]{Hirashitaetal2010}. The latter might be able to 1) trigger grain shattering, especially if the grains
reside in highly dense clumps \citepalias[with a clump lifetime the order of a few decades after they are crossed by the RS,][]{MartinezGonzalezetal2017} and 
2) restrict the motion of charged and large ($\sim 0.05$ $\mu$m) grains relative to the gas \citep{Fryetal2018}. Kinetic decoupling between gas and dust grains
is caused by their different inertial motions once they are processed by the RS, and also by the action of betatron acceleration of charged/large grains \citep{Bocchioetal2016}.
On the contrary, the gas-grain relative velocity is reduced by collisional and plasma drag deceleration \citep{DraineandSalpeter1979}. Grain acceleration 
within the cluster's volume might be effectively counterbalanced by the injection of the dense shocked stellar winds which increase the number of collisions and therefore the 
coupling of the gas and dust motions. If the grains are porous instead of solid spheres, then the larger grain area should also facilitate collisional coupling with 
the gas \citep{Jones2004}. Finally, as dust grains tend to be neutral at high gas temperatures \citep[$\gtrsim 2 \times 10^5$ K for silicate/graphite grains and $\gtrsim 3 \times 10^7$ K for Fe grains,][]{McKeeetal1987,Fryetal2018}, it is likely that 
betatron acceleration will not play a crucial role in setting apart the gas and dust motions. A suppressed kinetic gas-grain decoupling could imply that 
accounting for additional non-thermal destruction mechanisms might not make much difference, especially where thermal sputtering is already very efficient by itself 
(e.g. within the cluster's core).

\section{Summary and Concluding Remarks}
\label{sec:conclusions}

We have presented three-dimensional hydrodynamical simulations of dusty SNRs evolving in the steep density gradient established by shocked stellar 
winds in young massive clusters. In the model, dust grains are injected and advected with the SN ejecta. Thus, we have derived the fraction of 
SN dust mass which is injected into the circumcluster medium after shock-processing. We also explored the case of SNRs evolving 
in the diffuse ISM finding an excellent agreement with the results obtained by \citet{Bocchioetal2016} within the first $10^4$ years. In 
these cases, the log-normal grain size distribution results in $\sim 10\%$ more surviving dust mass in comparison to the standard 
\citetalias{MRN1977} distribution.

However, the hydrodynamical evolution of SNRs within massive clusters, where most massive stars reside, is radically different to that of isolated SNRs.
This evolution strongly depends on the interplay between the rapid adiabatic expansion and cooling after blowout, and the replenishment of 
mass and energy in the same region occupied by the SNR via shocked stellar winds \citep{Silichetal2017}. Thus, in many cases, the SN reverse 
shock crosses already thermalized ejecta and the fate of SN dust grains is sealed at early post-explosion times. The fraction of surviving SN 
dust mass is a strong function of the cluster's mass and the position of the SNe with respect to the cluster's center. These dependences are 
manifested as follows: 

\begin{enumerate}

\item In the event of well-centered SNe, as much as $30\%$ of the condensed dust mass survives the shock processing they encounter. Destruction 
is primarily induced by the interaction with shocked stellar winds rather than due to the crossing of the reverse shock. Given 
that a large fraction of the SN-condensed grains are efficiently heated to hundreds of K and subsequently destroyed, they should be observed 
transiently at NIR-MIR wavelengths \citep{MartinezGonzalezetal2016,MartinezGonzalezetal2017}. There are two sub-cases for well-centered 
explosions according to the cluster's mass:

\begin{enumerate}
\item For young stellar clusters with masses $\sim 10^6$ M$_{\odot}$, gas cooling induced by gas-grain collisions is very efficient as the grains 
are injected in multiple events. As originally shown by \citet{TenorioTagleetal2013,TenorioTagleetal2015}, this radically alters the 
thermodynamics of the star cluster wind.

\item For young stellar clusters with masses $\lesssim 10^5$ M$_{\odot}$, the average wind number density within the cluster is low 
and a larger fraction of the dust mass is able to stream out of the cluster (as much as $\sim 30\%$  in the present models).
\end{enumerate}

\item Off-centered SNe experience a blowout phase where the remnant elongates in the direction of decreasing wind density 
and protruding RT instabilities develop. As the SN ejecta cools down and lowers its density faster than in the case of an SN 
occurring in the diffuse ISM, thermal sputtering becomes less efficient and a non-negligible amount, up to $50\%$ in the present models, 
of the dust mass could be injected into the circumcluster medium.
\end{enumerate}


Given the assumed stellar density profile, which approximates the (deprojected) King stellar surface density profile 
\citep{King1962,Ninkovic1998}, a large fraction of the massive stars reside outside the core of the cluster, and thus, also a large 
fraction of SNe should experience a blowout phase.

\textit{We have found that a larger dust mass fraction is capable of surviving shock-processing in the case of SNe occurring in SSCs than in 
the case of isolated spherical SNRs. Not only that, we predict that clustered SN explosions will cause a net increase in the amount of 
dust in the surroundings of young massive stellar clusters} after having survived shock-processing due to the cluster wind, the passage of the 
reverse and secondary forward shocks and the crossing of subsequent SN forward shocks. As the large quantities of dust observed in galaxies at 
extreme redshifts require both, efficient dust formation and efficient dust survival rates, our model might serve to alleviate the 
tension between observational and theoretical expectations, the so-called ``dust budget problem'' \citep{Ferraraetal2016}.

The grains which stream out of the cluster will face milder conditions: far from the bulk of the starlight and the hot environment within                                                                                 
the cluster. These grains should accumulate in the circumcluster medium and manifest persistently at MIR-FIR wavelengths. We plan to quantify this emission 
and compare it with observations of nearby massive clusters and blue compact dwarf galaxies. We have restricted our analysis to the evolution of multiple SNRs 
in clusters as massive as $10^6$ M$_{\odot}$. Clusters which are even more massive have a higher SN rate and a larger gravitational potential; under 
certain conditions (e.g. enhanced radiative cooling), they might be able to retain a non-negligible fraction of SN ejecta, leading 
to enhanced Fe abundances in secondary stellar generations. Addressing this scenario is also in our plans.

We finally note that the fast-moving dusty ejecta in the blowout phase resemble various aspects of volcanic mushroom clouds upwardly 
expanding into a stratified atmosphere after violent super-eruptions \citep[see the hydrodynamical simulations of such events by][]{Costaetal2018}. 
In that case, the pyroclasts\footnote{The term pyroclastic derives from the greek roots {\it pyros}, meaning ``fire'' and 
{\it klastos}, meaning ``broken in pieces''.}, i.e. ashes and cinders, carried by the unstable cloud are wind-driven and eventually 
accumulate into continent-size regions.

\acknowledgments

This study has been supported by project 17-06217Y of the Czech Science Foundation and by the institutional project RVO:67985815,
by CONACYT - M\'exico, grant 167169, and by the Spanish Ministry of Science and Innovation for the ESTALLIDOS collaboration (grants 
AYA2013-47742-C4-2-P estallidos5 and AYA2016-79724-C4-2-P estallidos6). The computational time was provided by the Czech 
Ministry of Education, Youth, and Sports from the Large Infrastructures for Research, Experimental Development, and Innovations 
project "IT4Innovations National Supercomputing Center - LM2015070". The software used in this work was in part developed by 
the DOE NNSA-ASC OASCR Flash Center at the University of Chicago. The authors thank the anonymous Referee for a careful 
reading and helpful suggestions which greatly improved the paper. S.M.G. acknowledges the hospitality of the IAC during his 
visit in 2017 and support of the Erasmus+ programme of the European Union under grant number 2017-1-CZ01-KA203-035562.

\appendix

\label{app}

\section{Log-normal grain size distribution}
\label{app:A1}

\begin{figure}
\epsscale{0.75}
\plotone{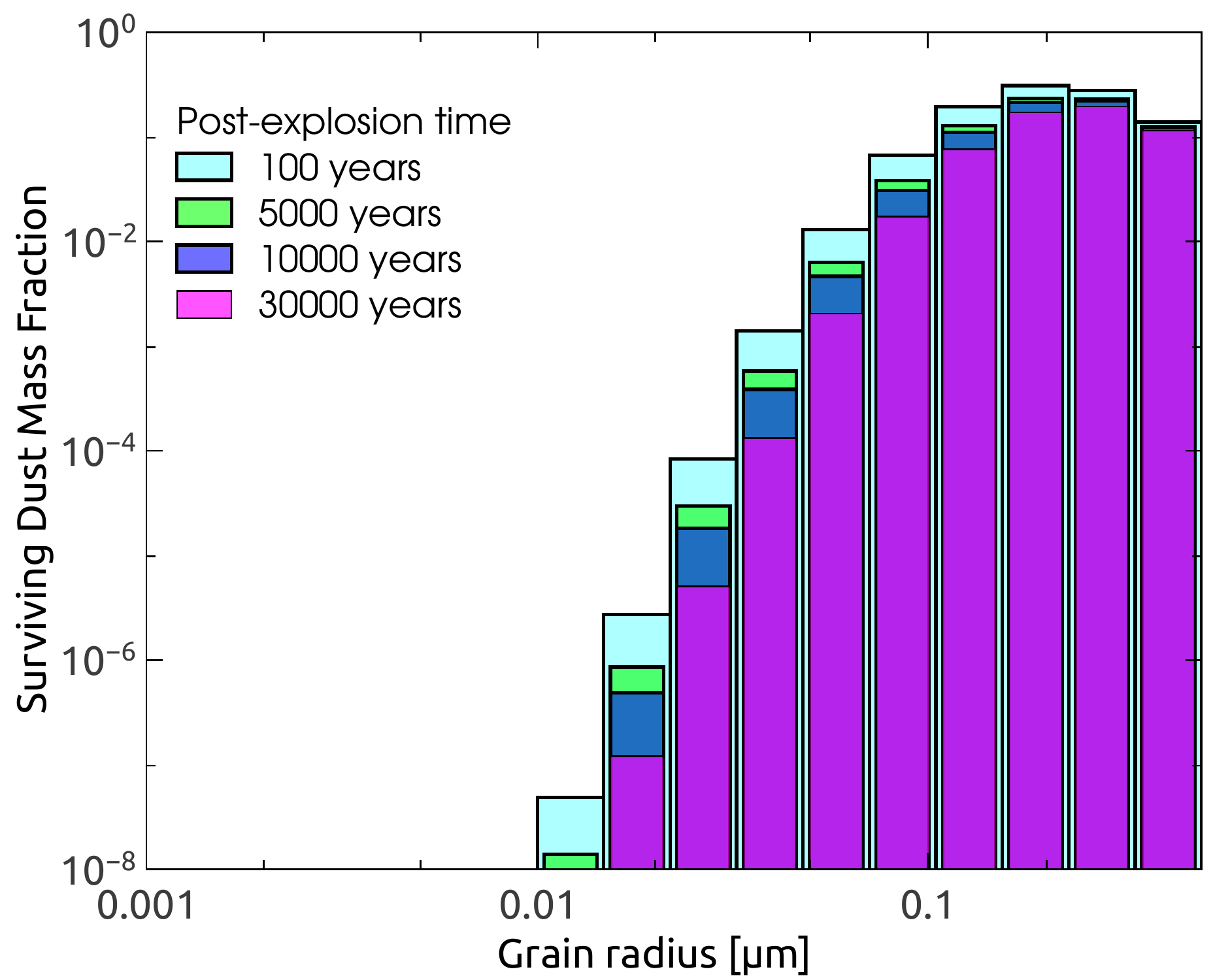}
\caption{Dust mass evolution per grain size bin in our stratified run ($\omega=1$) with an initial log-normal grain size 
distribution. The histograms depicts the distribution of dust mass per size bin at different post-explosion times 
(light blue: 100 years, green: 5000 years, dark blue: 10000 years, violet: 30000 years).}
\label{fig:A1}
\end{figure}

Since type II-P SNe represent the majority of core-collapse SNe \citep{Sukhboldetal2016}, and form preferentially large 
grains \citep{Kozasaetal2009}, it is interesting to study the case where the initial grain size distribution is strongly 
dominated by large grains instead of the classical \citetalias{MRN1977} grain size distribution, which is more appropriate 
to the ISM dust. We have explored the same SN model as in our stratified fiducial run ($\omega=1$) but with a particular case 
of a log-normal grain size distribution \citep[often assumed in characterizing the dust yield of SNe; e.g.][]{Bocchioetal2016} 
of the form $\sim a^{-1} \exp \{-0.5 [\log(a/a_{0})/\sigma]^2\}$, with $a_{0}=0.1$ $\mu$m and $\sigma=0.7$ and lower and upper 
limits $a_{min}=0.01$ $\mu$m and $a_{max}=0.5$ $\mu$m, respectively. We have modified the logarithmic binning and the dust mass 
corresponding to each size bin accordingly. In Figure \ref{fig:A1}, similarly to \ref{fig:2}, we show the dust mass evolution 
per grain size bin at four different post-explosion times (100, 5000, 10000 and 30000 years). In Figure \ref{fig:A2} the blue 
dashed-double-dotted curve depicts the time evolution of the dust mass in comparison to the fiducial run with the 
\citetalias{MRN1977} (black solid line). The selected particular case of the log-normal grain size distribution results in the 
survival of an additional $\sim 10\%$ of the total dust mass with respect to the \citetalias{MRN1977} case.

\section{Steeper ejecta density profiles}
\label{app:A2}

\begin{figure}
\epsscale{0.75}
\plotone{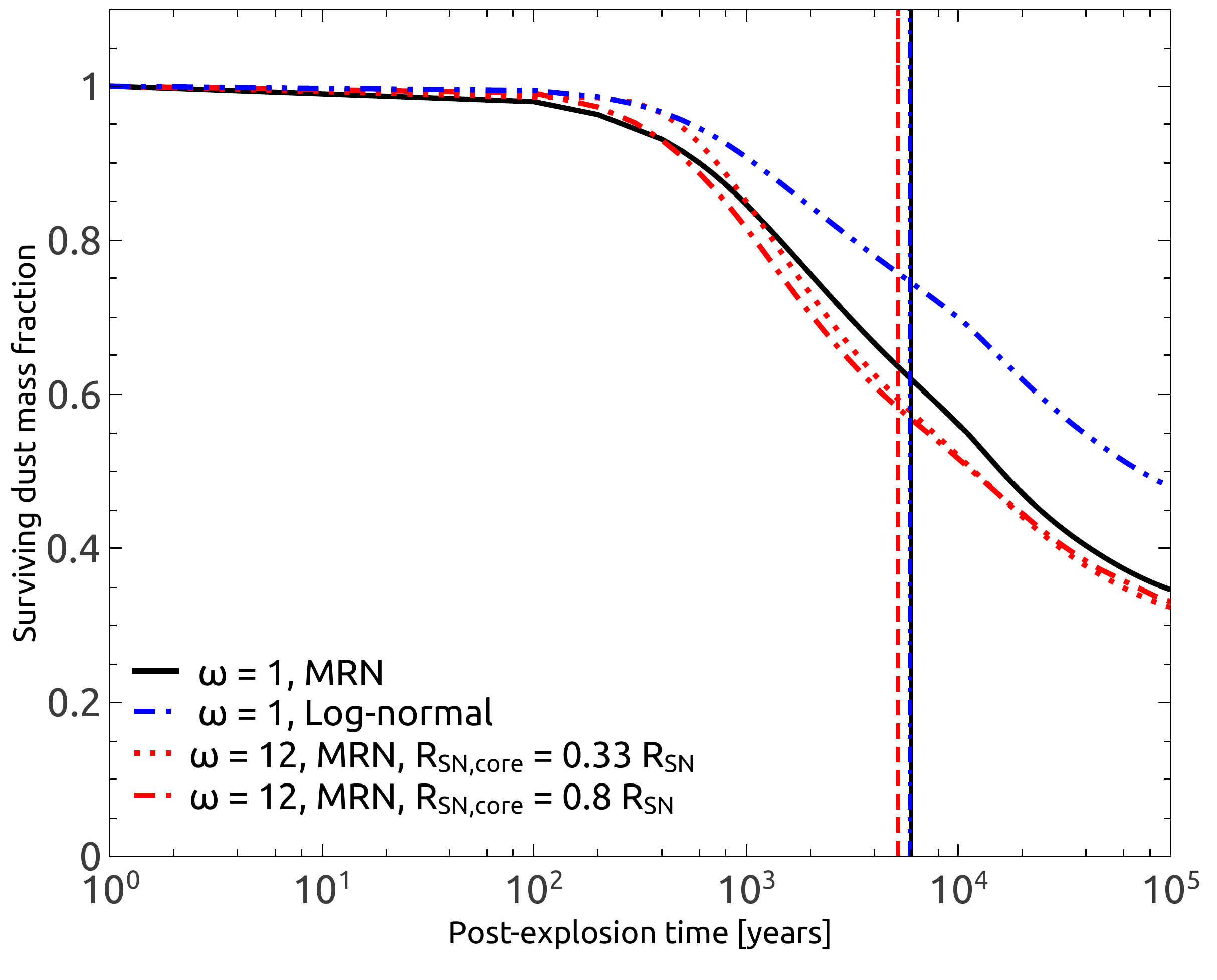}
\caption{Same as in Figure \ref{fig:3} but for the ``log-normal'' and ``steep ejecta profile'' cases as compared to the fiducial
stratified model.}
\label{fig:A2}
\end{figure}

It is also interesting to test cases with even steeper ejecta density profiles than those studied in Section 
\ref{sec:isolated}. Therefore, we have performed two additional cases of isolated explosions with the same input parameters as 
in our fiducial runs, but with ejecta density profiles composed of a flat inner core of initial size $R_{SN,core}$, and an 
outer region which falls off as a power-law \citep[$\rho_{ej}\sim d^{-\omega}$, see equations A1-A5 in][]{TangChevalier2017}. 
Both cases consider an outer steep region with $\omega=12$, while the flat core is characterized by $R_{SN,core}=0.33 R_{SN}$ 
in the first case, and $R_{SN,core}=0.8 R_{SN}$ (enclosing $\sim 3/4$ of the ejecta mass) in the second case. Indeed, 
\citet{Chevalier1982} modeled the ejecta profile of type I SNe using $\omega=7$ only for the outer $3/7$ (by mass) of the 
ejecta and pointed out that the outer ejecta in core-collapse SNe would be steeper and the mass contained outside their flat 
inner cores would be considerably smaller than in type I SNe. Moreover, \citet{LuoandMcCray1991} assumed $\omega=9$ for the 
outer $3$ M$_{\odot}$ of the ejecta, whereas $6$ M$_{\odot}$ were located in the inner flat region of the ejecta of SN 1987A. 
The implication that most of the ejecta is contained within $R_{SN,core}$ is that the results obtained for our fiducial runs 
in Section \ref{sec:isolated} will only differ for the short time \citep[typically less than 1000 years, see equation 25 in ][]{Chevalier1982} 
the RS takes to reach the inner flat core, especially if this time is comparable to the 
characteristic timescale for dust sputtering. In Figure \ref{fig:A2} we show that assuming an outer steep region with 
$\omega=12$ does not significantly affect the amount of surviving dust when compared to the shallow stratified fiducial run
with $\omega=1$. Also note that reducing the size of the inner flat core does not change the results significantly.

\bibliographystyle{apj}
\bibliography{Infrared}

\end{document}